%% file: lsa.tex
\documentclass[copyright,creativecommons]{eptcs}
\usepackage{graphicx}
\usepackage{subfigure}
\usepackage{latexsym}

\usepackage{breakurl}        

\title{Re-verification of a Lip Synchronization Protocol using Robust Reachability}
\author{Piotr Kordy  \qquad \quad \quad$\;$ Rom Langerak\qquad \quad \quad$\;$$\;$Jan Willem Polderman\quad$\;$
    \email{kordy@cs.utwente.nl \qquad langerak@cs.utwente.nl \qquad j.w.polderman@math.utwente.nl}
    \institute{Formal Methods and Tools\qquad$\;$ Formal Methods and Tools\qquad Mathematical Systems and Control Theory}
    \institute{University of Twente, Drienerlolaan 5,  7522 NB Enschede, The Netherlands}
}

\begin{document}
\maketitle
\input{macros}
\begin{abstract}
The timed automata formalism is an important model for specifying and analysing real-time 
systems.  Robustness is the correctness of the model in the presence of small drifts on 
clocks or imprecision in testing guards. A symbolic algorithm for the analysis of the 
robustness of timed automata has been implemented.  In this paper, we re-analyse an 
industrial case lip synchronization protocol using the new robust reachability algorithm.
This lip synchronization protocol is an interesting case because timing aspects are crucial
for the correctness of the protocol. Several versions of the model are considered: with an
ideal video stream, with anchored jitter,  and with non-anchored jitter.
\end{abstract}
\section{Introduction}
Timed automata \cite{AD94} is a widely used and successful formalism to
analyse real-time systems.
Timed automata are automata extended by clock variables that can be
tested and reset. Numerous
real-time systems have been specified and analysed by the tool UPPAAL
\cite{BY03,uppaal}
and the approach can be said to be mature and industrially applicable.

However, if we want to implement a system then robustness becomes an issue.
We need to know if the system
is resilient with respect to small perturbations. Timed automata in
its original form may be crucially dependent on perfect precision
of the clocks. Therefore, several publications have suggested alternative
semantics of
timed automata that take into account
perturbations. In particular, the skewed clocks automata from
\cite{Alur93} can have arbitrary rates for clocks, the
``tube languages'' from \cite{GHJ97,HR00} deal with open sets of
trajectories, ``perturbed'' timed automata
from \cite{Puri2000} are subjected to an infinitesimal noise,
and the implementable timed automata from
\cite{Wulf2004,Wulf2008} should be implementable using discrete clocks.

Semantics can only be said to be successful if it can be applied in
practice. In this paper we are interested in the work that was initiated 
by Puri \cite{Puri1998}. He considered drifting clocks and
showed that timed automata models are not robust with respect to
safety properties, meaning that a model proven to be safe under the standard ideal semantics might not
be safe even if clocks drift by an arbitrarily small amount. The
region based algorithm has been proposed
to calculate set of states that are reachable for \emph{any} clock
drift. Puri's approach
has been  extended by the introduction of the concept of stable zone
\cite{Daws2006}, which
made it possible to implement
an efficient algorithm that can be used in practice.

To check the new performance of the new algorithm the best way is to
apply it to an
industrial case study.
On the UPPAAL homepage \cite{uppaal} a number of case studies can be
found
in which the UPPAAL tool has been applied.
We investigated several of them and we have chosen the case study where a
lip
synchronisation algorithm is analysed.
This algorithm is used to synchronize multiple information streams
sent over a communication network,
in this case audio and video streams of a multimedia application. We
chose this case study mainly because timing
is an important aspect of synchronisation and therefore this algorithm
can be expected to be
sensitive to  small disturbances of the
clock drift.

\paragraph{Structure of the paper}
The rest of the paper is organised as follows. Section \ref{sec:lsp} 
introduces the lip synchronisation problem. Section \ref{sec:mf} 
presents the modelling formalism and tool used in the analysis of
the lip synchronisation protocol. Section \ref{sec:model} provides a 
description of a model of the lip synchronisation protocol. Section 
\ref{sec:verification} presents the verification results and 
Section \ref{sec:conclusions} gives a concluding discussion.

\section{Lip Synchronisation Problem}
\label{sec:lsp}
The problem of lip synchronisation has been present in the literature \cite{StefaniHH92,AtesBSS96}. 
Here we present it briefly, for more detailed description look in
\cite{Bowman98}. In this paper, we consider the problem of 
synchronising of audio and video streams. We consider a scenario when audio and video
are transmitted as separate streams that need to be synchronised at the sink.

\begin{figure}[htpb]
\centering
   \includegraphics[height=0.6\textwidth]{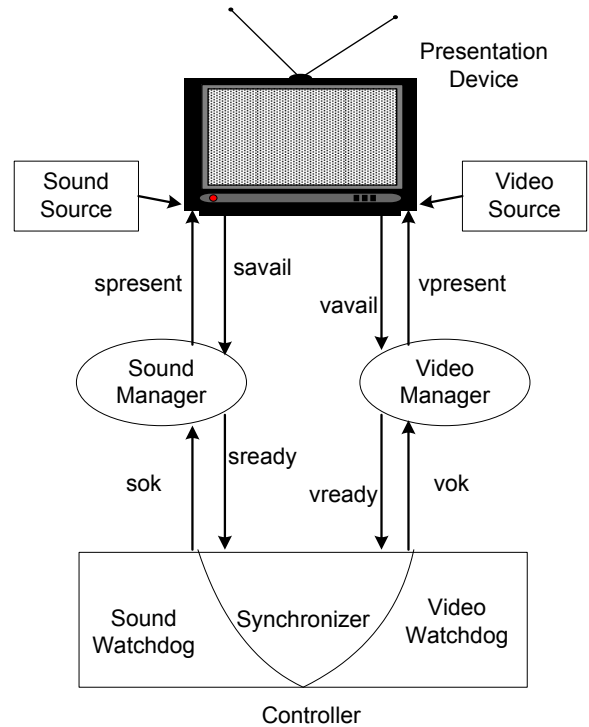}
\caption{Overview of the structure of the lip synchronisation system}
\label{fig:lls}
\end{figure}

The overview of the basic configuration can be seen in Figure \ref{fig:lls}.
There are two stream sources: one for sound and one for video. These streams arrive at a presentation 
device. We need to ensure that both streams play synchronised within certain level of tolerance. This is the 
problem that lip synchronisation protocol addresses.

The protocol is implemented using several components: \emph{sound} and \emph{video} 
managers, and a \emph{controller}. Communication between components is done using signals. 
When presentation device receives sound packet it sends a \emph{savail} 
\footnote{names are usually prefixed with s for sound and v for video} signal to the 
\emph{Sound Manager}. At an appropriate moment the Sound Manager sends \emph{spresent} signal
to the Presentation Device to indicate that the packet should be played. The
\emph{Video Manager} has similar behaviour and uses signals \emph{vavail} and
\emph{vpresent}. The \emph{Controller} contains the main body of the protocol.
It receives signals \emph{sready} and \emph{vready} from the managers. The signals
indicate that sound and/or video packets can be presented. The Controller decides if it is 
the correct time to play the packet. Confirmation is done using \emph{sok} and \emph{vok}
signals, respectively. If it is not possible to synchronise, the Controller signals an error and
enters an error state.

The requirements for acceptable synchronisation between the two streams are the following:
\begin{itemize}
	\item The time granularity is 1 millisecond.
	\item A sound packet is presented every 30 milliseconds and no jitter is allowed.
	\item Optimally, a video packet should be presented every 40 milliseconds. However, 
        we allow some margin of error:
\begin{itemize}
	\item video frames may precede sound frames up to 15 milliseconds and may lag 
        up to 150 milliseconds.
	\item we allow a 5ms jitter that is, a video package may be no less than 35 ms late 
        and no more than 45 ms away from ideal presentation time (Anchored Jitter) or 
        from previous packet (Non-anchored Jitter)
\end{itemize}
\end{itemize}

\section{Modelling Formalism} 
\label{sec:mf}
\subsection{Timed Automata}
For our purposes we use the existing timed automata model from \cite{Bowman98}. The modelling 
formalism is based on a network of timed automata. The network of timed automata consists of the parallel 
composition of a number of timed automata, and a configuration. A timed automaton is an 
automaton consisting of locations and edges that are extended with real valued variables 
called clocks. 

Edges and locations are labelled. The labels of an edge may consists of several optional components: a guard, a 
synchronisation label, a set of clocks to reset, and assignments to integer variables.
The guard on clocks and/or on data variables expresses under which conditions we are allowed to take a transition. If there is no guard, the condition is interpreted 
as true. Because later we consider a so called robust semantics, we limit guards on clocks to the 
form : $\x\leq c$ or $\x \geq c$ where $\x$ is a real valued clock and $c$ is natural 
number. We also allow a formula that is a combination of the above terms using logical \emph{and}.
When we take a transition we may perform a synchronisation. The synchronisation label must
be synchronised with its counter part. The synchronisation rules are similar as in CSS
\cite{Milner1995}. When there is no synchronisation, a label is interpreted as an internal 
action (similar to $\tau$-actions).

The labels of locations consist of the name of a location, an optional invariant, and can be 
marked as \emph{committed} or \emph{urgent}. An invariant is a constraint on clocks, 
indicating how long we can stay in a location. It is similar to guard but only upper
bound constraints are allowed. When a location is marked as committed, we have 
to leave this location without any delay or any interleaving actions. This is useful
to ensure atomicity of sequence of transitions. When a location is urgent time is 
not allowed to pass in that location.

The configuration consists of the names of timed automata composing the system, global 
variables, and channels. The synchronisation happens through channels and synchronisation
labels are names of channels. Channels can be \emph{urgent}. When a channel is urgent,
we have to take that transition as soon as possible that is without delay.
There can be no guards on edge with urgent channel.

The state of timed automaton is of the form $\left(\bar{\q},\va\right)$ where $\bar{\q}$ 
is a control vector and $\va$ is the clock valuation. The control vector shows the current
location for each timed automata in the network and the clock valuation indicates value
of each clock and integer variables. The initial state consists of initial
locations for each timed automaton and all clocks and variables equal to value $0$.
From a state it is possible to take two types of transitions: \emph{delay} and 
\emph{edge} transition. When we take a delay transition, all clocks are progressing
at the same speed within the values allowed by the location invariants. An edge transition can 
be internal or a synchronisation. An internal transition can occur when the network is at 
the location in which it can take an edge with no synchronisation label. The guard must 
be satisfied by the current clock valuations. A synchronisation transition occurs when 
two edges can synchronise over complementary synchronisation labels. Guards of both 
edges must be satisfied.

\subsection{Model Checking}
The continuous time leads to infinitely many states. Fortunately, as noted in \cite{AD94} 
similar states can be grouped into  \emph{regions}. However region automaton is not 
the most efficient representation of the state space of a timed automaton. It suffers from 
a combinatorial state explosion, which is dependent on the size of the constants used. 
\emph{Zones} are used as a more efficient representation of the state space 
\cite{Dill1990,Henzinger1994,Yannakakis1993} as they represent the state 
space in a more aggregated way.
In the tool UPPAAL \cite{BY03} more effective zone based algorithm is used.

The UPPAAL tool is able to check for reachability properties. Those properties are of 
the form:
\[
\varphi::=\forall \Diamond \beta \;|\; \exists \Box \beta \quad\quad\quad\quad%
\beta::= a \;|\; \beta_1 \wedge \beta_2 \;|\;\neg \beta \;|\;\beta_1 \Rightarrow \beta_2
\]
where $a$ is an atomic formula being either an atomic clock (or data) constraint or a component location $(A_i\;\; \mbox{at}\;\;l )$.
Atomic clock (data) constraints are integer bounds on individual clock (data) variables 
(e.g. $1 \leq x \leq 3)$.

Intuitively for $\forall \Box \beta$ to be satisfied all reachable states must satisfy 
$\beta$. For $\exists \Diamond \beta$ to be satisfied some reachable state must satisfy $\beta$.

\subsection{Robustness Problem}

Clocks in a timed automata network are synchronous. Puri \cite{Puri2000} has shown that this 
assumption is not robust to even infinitely small clock drifts. In short, it means that we 
can reach states that are not reachable in normal semantics for \emph{any} value of 
the clock drift. He proposed new 
reachability semantics for timed automata and we will call it the \emph{robust semantics}.
The idea is to have a parametrised reachability, where a parameter defines how much clocks are allowed to drift. 
In normal reachability, when time progresses by $t$ time units, the new clock valuation is 
$\va+t$. When we allow clock drift parametrised by $\eps$, the new clock valuation can have
$\eps$ small differences between clocks. Formally
\[
\va'\left(\x_i\right) - \va\left(x_i\right) \in \left[\left(1-\eps\right)t, \left(1+\eps\right)t\right],\mbox{ for }i= 1,\ldots,n
\]
where $n$ is number of clocks, $\va'$ is a valuation after the time transition and $\va$ 
is a clock valuation before the time transition. Let $\ReachE{\s_0}$ denote the reachable set of states from the initial state $\s_0$ in the robust semantics. Unfortunately parametric 
model checking of timed automata with three clocks and only one parameter is known to be 
undecidable \cite{Alur1993,WongToi1999}. What Puri proposes is the reachability 
when the drift $\eps$ is infinitely small. Formally
\[
\StarReachE{\s_0}=\bigcap_{\eps>0}\ReachE{\s_0}
\]
Puri shows that calculating $\StarReachE{\s_0}$ is decidable. He proposes a region based 
algorithm.

\begin{figure}%
\centering
\subfigure[First.][A timed automaton]{
	  \includegraphics[width=0.7\textwidth]{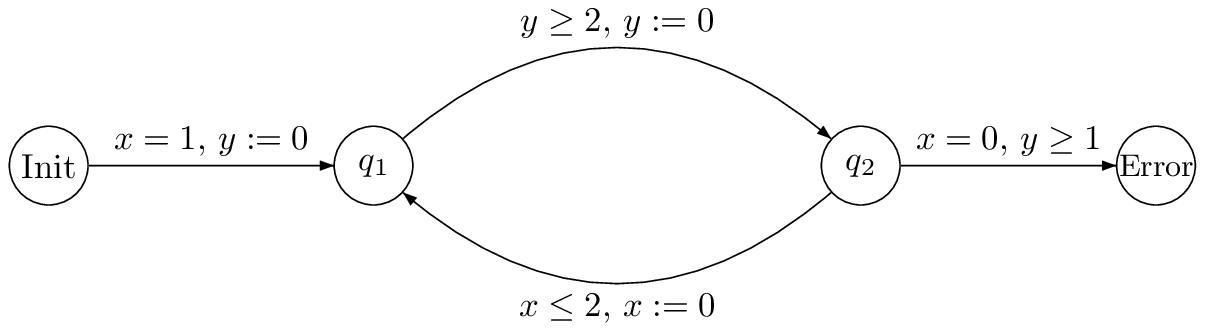}
	  \label{fig:examplePuri}
}\\
\subfigure[Second figure.][Reachable set of states in the normal semantics]
{
  \includegraphics[width=0.7\textwidth]{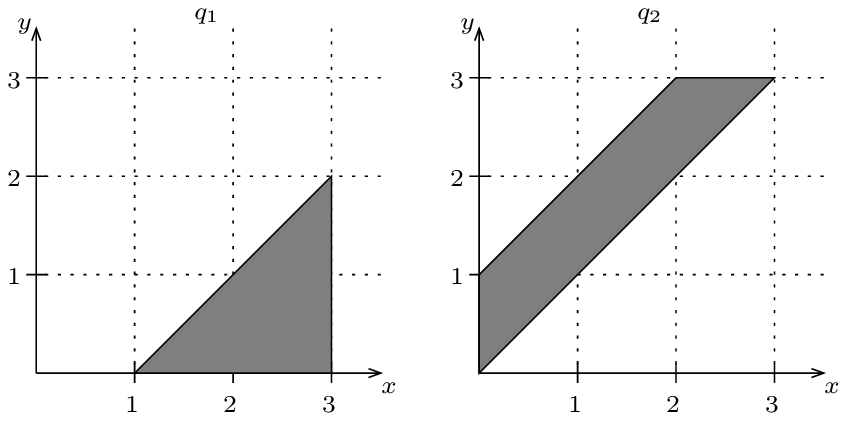}
  \label{fig:reachGraphA}
}\\
\subfigure[Third.][Reachable set of states in the robust semantics]{  
   \includegraphics[width=0.7\textwidth]{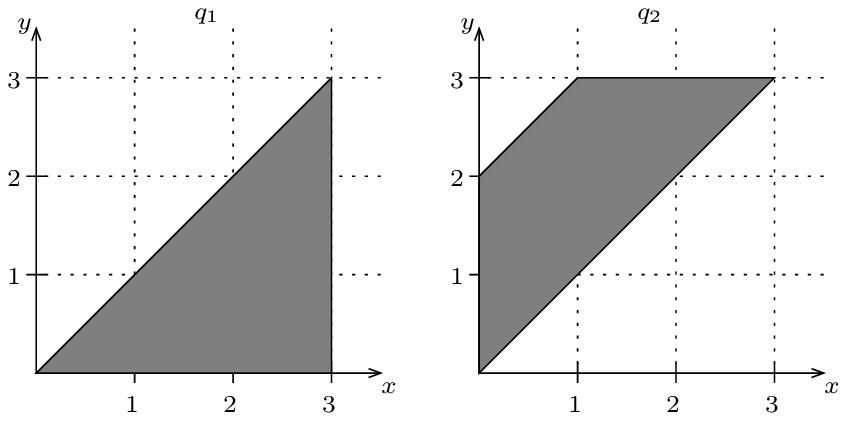}
   \label{fig:reachGraphB}
}
\caption{Example of timed automaton with different reachability under normal and robust semantics.}
\label{fig:3figs}
\end{figure}

It may seem that $\StarReachE{\s_0}$ and $\Reach{\s_0}$ are the same since $\eps$ is 
small, but this is not the case.
Consider the example shown in Figure \ref{fig:examplePuri}. This timed automaton 
has two clocks $\x$ and $\y$ and four locations. From location Init we can only go to 
location $\q_1$ and the value of the clocks will be $\x=1$ and $\y=0$. In the precise 
semantics, following the cycle between locations $\q_1$ and $\q_2$, we will get the 
reachable set of states  $\Reach{\Sem}{\s_0}$ depicted in Figure \ref{fig:reachGraphA}. 
We want to avoid Err location. The Err location is not reachable for both $\alpha=2$ and 
$\alpha=3$.

Now we consider the case for the robust semantics. Let $\e_1$ be edge from location $\q_1$ 
to $\q_2$ and $\e_2$ edge from location $\q_2$ to $\q_1$. Notice that for any 
$0\leq\beta\leq 1$ the following sequence of transitions is possible: 
$(\q_1;\x=\beta,\y=0)\stepE{2-\beta}\stepE{\e_1}(\q_2;x=0,\y=2-\beta+\eps)%
\stepE{\beta-\eps}\stepE{\e_2}(\q_1;x=\beta-\eps,\y=0)$. Hence, if we cycle sufficient 
number of times for any $\eps>0$ we can reach state $\left(\q_1;x=0,y=0\right)$ which 
is not reachable in the normal semantics $\Sem$. Thus, the Err location is robustly reachable 
for $\alpha=2$. This shows that the normal semantics is not robust with respect to small 
clock perturbations. Even small changes in the clock drift may lead to a dramatic change in 
the behaviour of a system. We avoid the Err location only in the normal semantics, but not in 
the robust semantics. We say that such safety property is non-robustly satisfied. If a
system has non-robustly satisfied property it is not implementable because its 
correctness depends on the mathematical idealization of the normal semantics.

\subsection{Verification Tool}
Puri proposed an algorithm to calculate $\StarReachE{\s_0}$ that is based on regions. 
Basically the algorithm finds regions that are on the cycle in the region graph. Such 
regions have the property that we can drift form any point to any other point in that 
region in robust semantics. When a part of such a region is encountered in the search, the 
whole region is added to the reachable set of states. In \cite{Daws2006} the 
notion of a \emph{stable zone} has been introduced. Basically the stable zone has the same property as the region on the cycle that is, we can drift from any point in the stable to any other point in that stable zone. Thus, this
is a good starting point for the zone based algorithm to calculate robust reachability. 
Together with the Aalborg University the prototype tool is being developed based on UPPAAL
$4.1.1$ and it is used in this paper.

\section{The Model}
\label{sec:model}
In this section UPPAAL model is presented. It follows the specification given in \cite{Bowman98} which in turn was derived from the specification given in LOTOS \cite{Regan93}. 
The model represents the specification of the video and sound managers and the 
synchroniser from Figure \ref{fig:lls}.
\begin{figure}[htbp]%
\centering
\parbox{1\textwidth}{%
\includegraphics[scale=0.7]{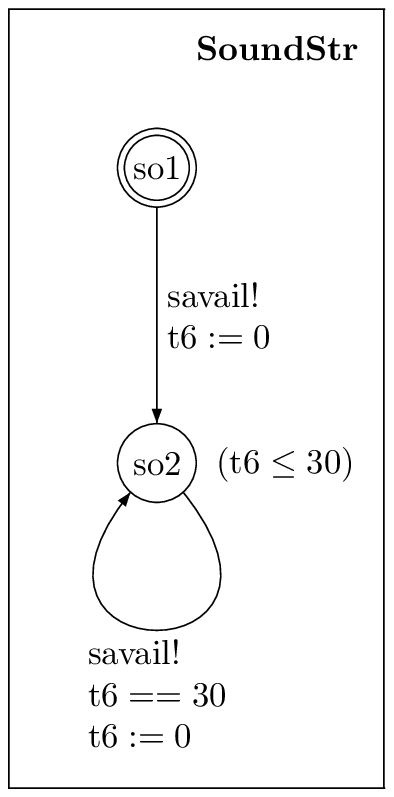}
\includegraphics[scale=0.7]{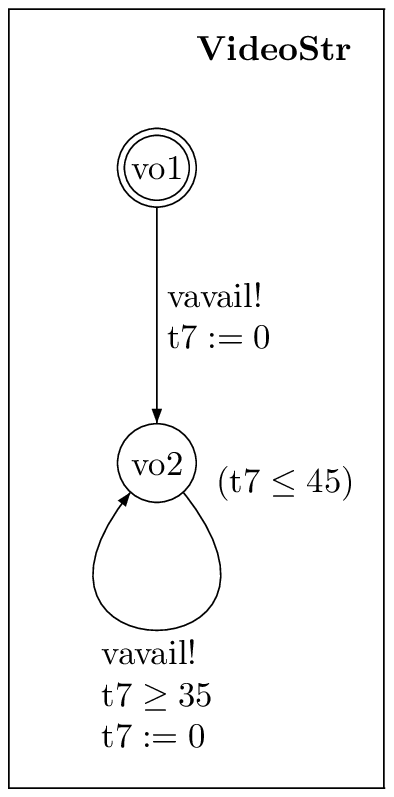}
\includegraphics[scale=0.7]{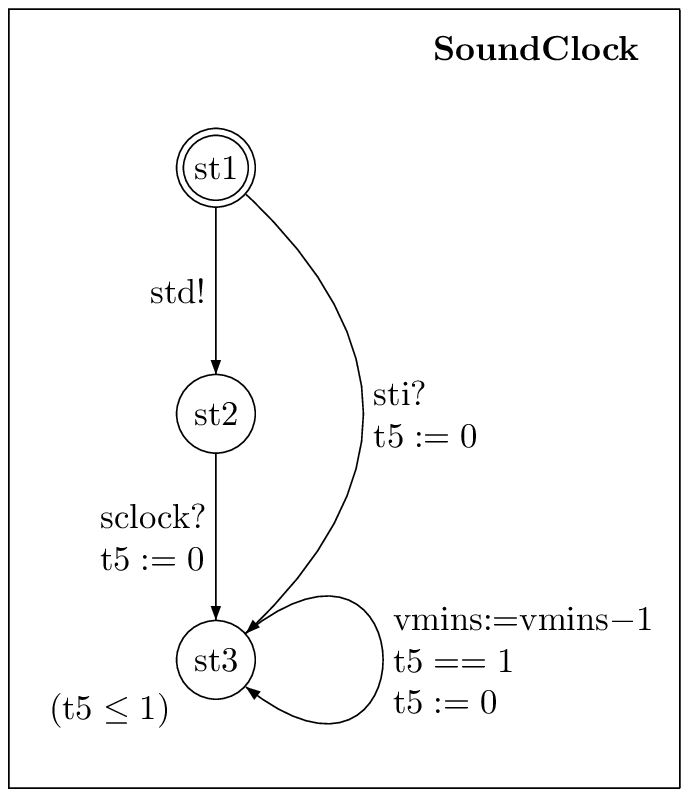}
\includegraphics[scale=0.7]{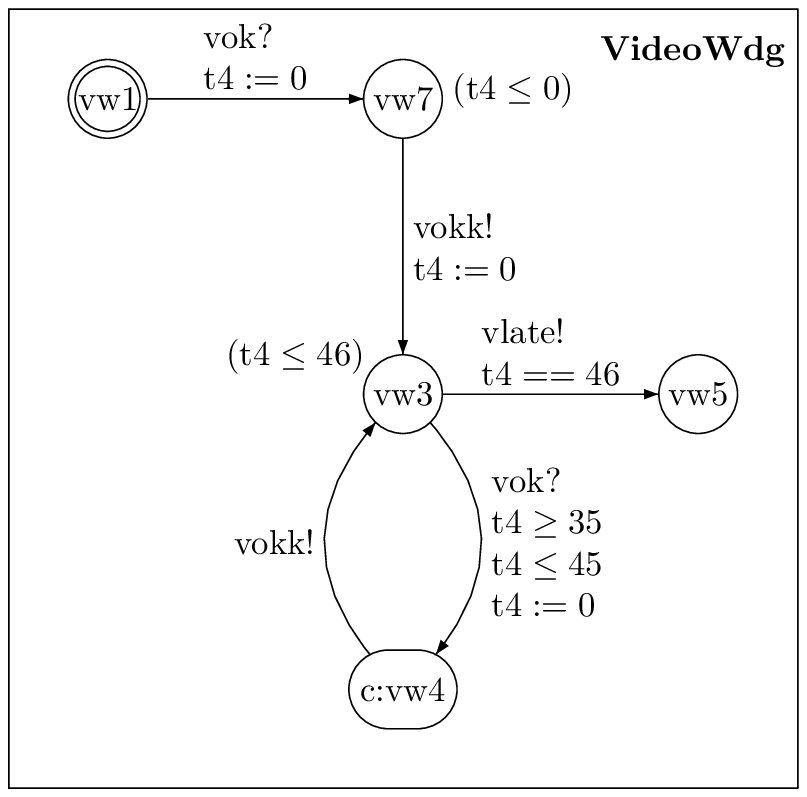}%
}
\vspace{1pt}
\\
\parbox{1\textwidth}{
\includegraphics[scale=0.7]{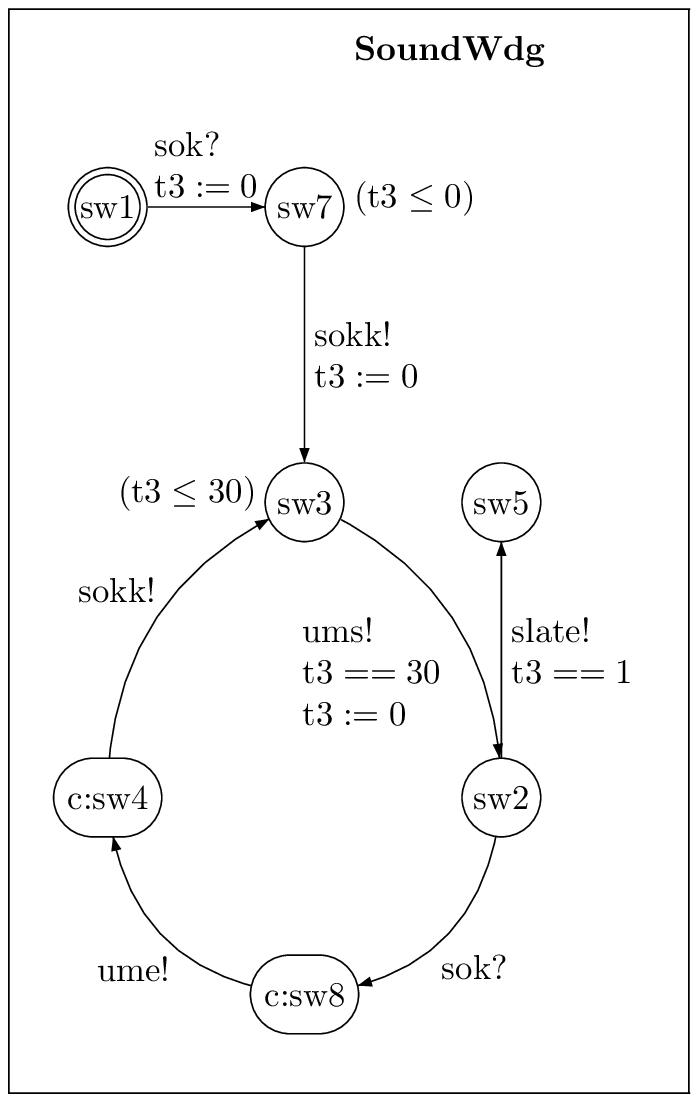}
\includegraphics[scale=0.7]{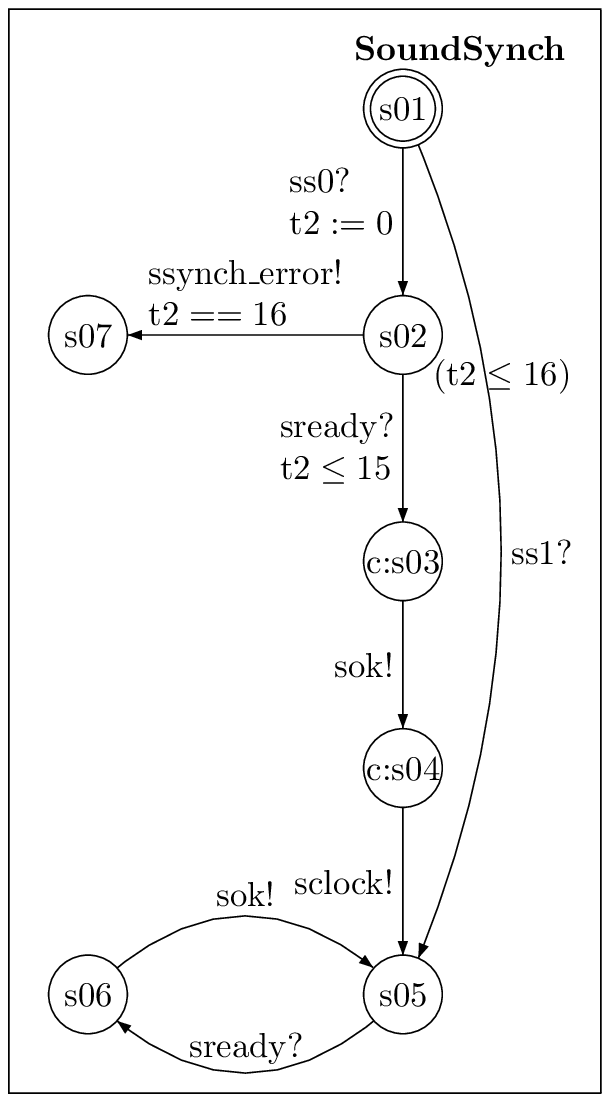}
\includegraphics[scale=0.7]{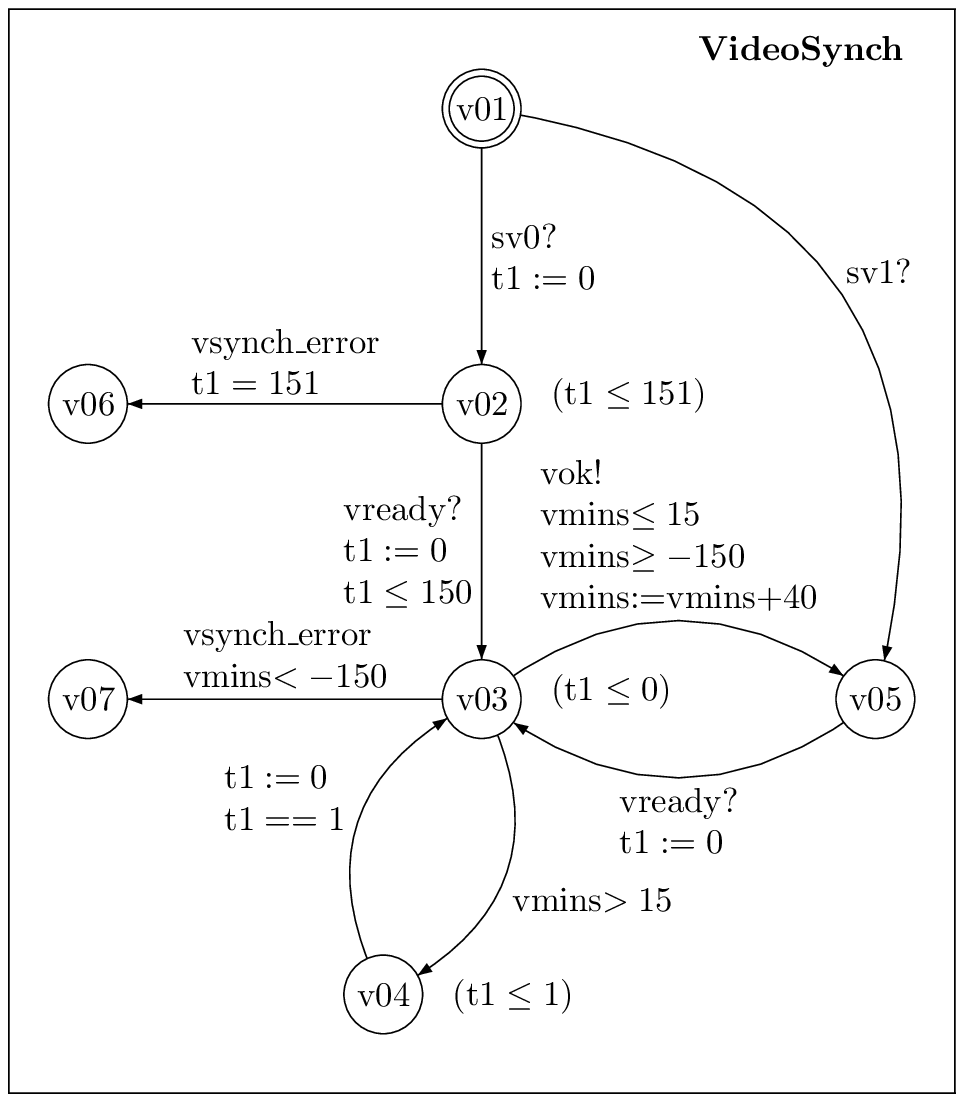}%
}
\vspace{1pt}
\\
\parbox{1\textwidth}{
\includegraphics[scale=0.7]{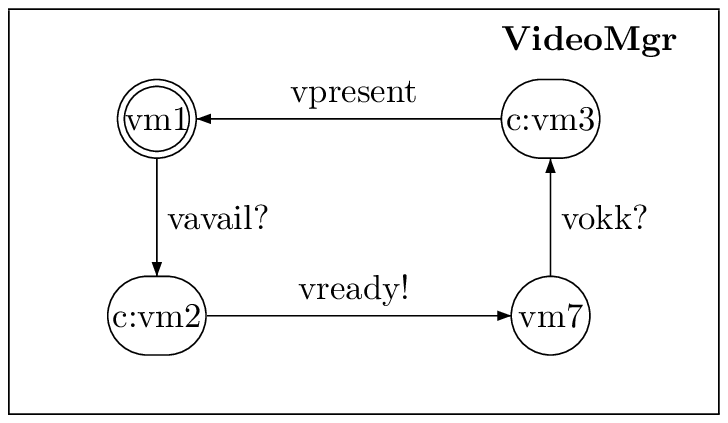}
\includegraphics[scale=0.7]{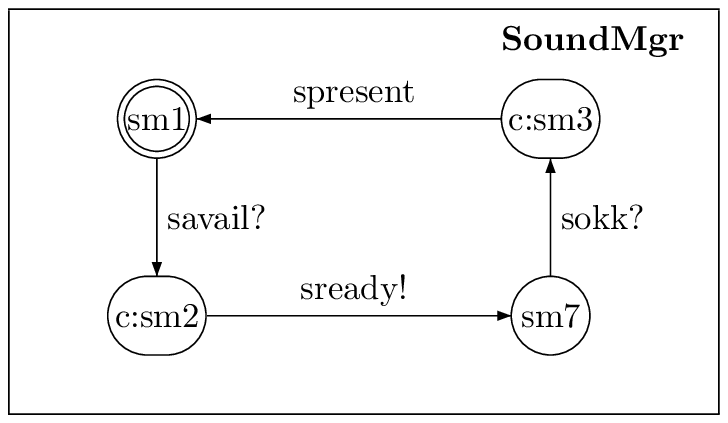}
\includegraphics[scale=0.7]{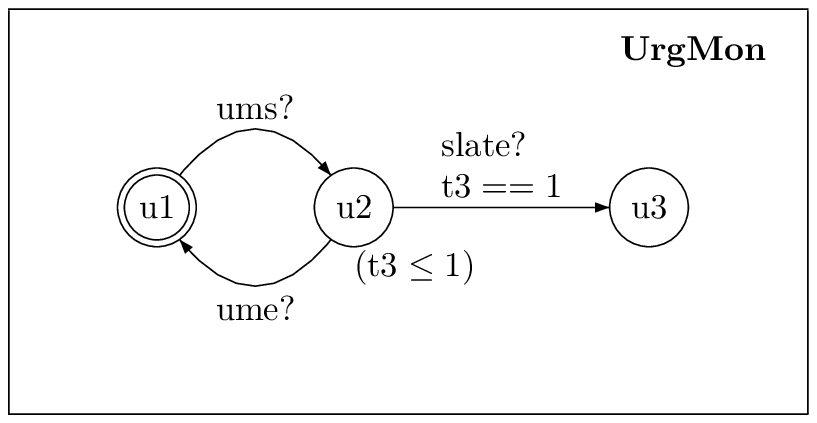}%
}
\vspace{1pt}
\\
\parbox{1\textwidth}{
\includegraphics[scale=0.7]{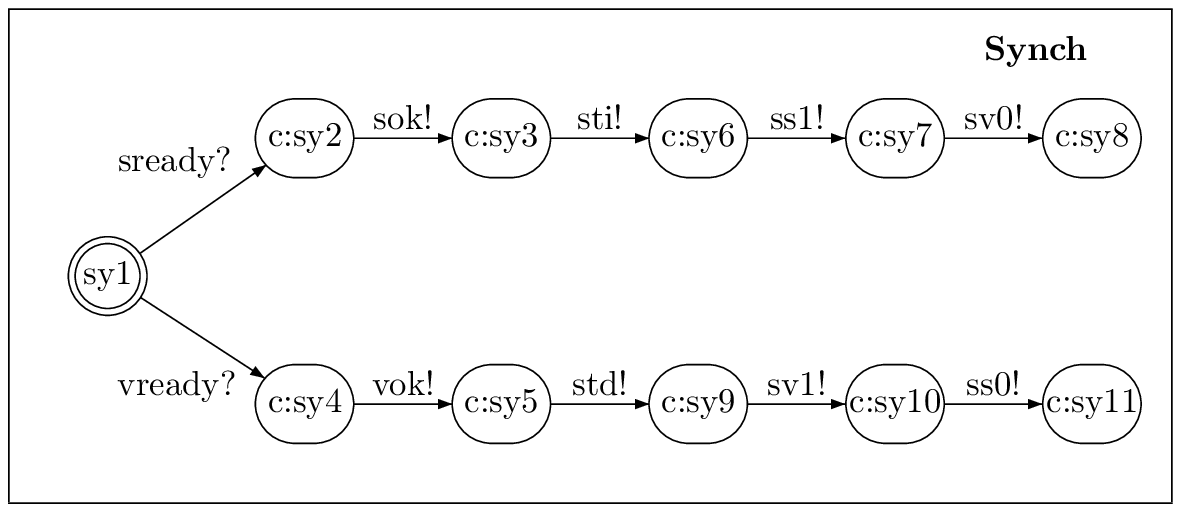}
\includegraphics[scale=0.7]{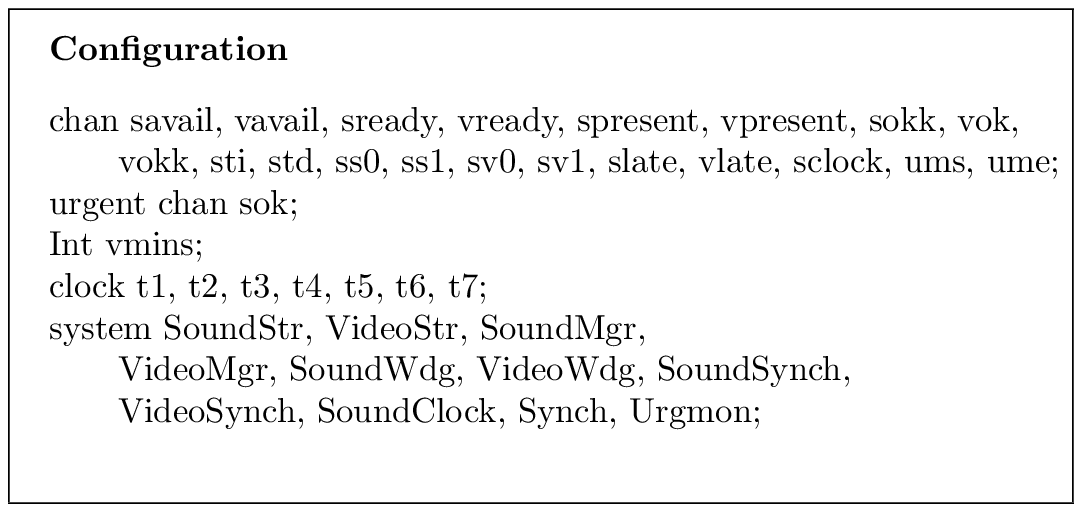}%
}
\caption{A model of Lip Synchronisation Protocol}
\label{fig:model}
\end{figure}

The model is shown in Figure \ref{fig:model}. Sound and Video Managers are modelled by 
automata \emph{VideoMgr}, \emph{SoundMgr}, \emph{VideoWdg}, \emph{SoundWdg} and 
\emph{UrgMon}. The Synchronizer consists of \emph{Synch}, \emph{VideoSynch}, 
\emph{SoundSynch} and \emph{SoundClock}. The external environment that is the sound and 
video streams are modelled by \emph{VideoStr} and \emph{SoundStr}. We briefly discuss the 
components.

\paragraph{The stream managers}
Both stream managers are quite simple. After receiving a signal \emph{vavail} or 
\emph{savail} indicating that the video or sound packet is available, they forward the signal 
immediately to the synchroniser using \emph{vready} and \emph{sready}. The immediacy 
is ensured by marking the locations \emph{vm2} and \emph{sm2} as committed. Next the  
manager waits for a confirmation from the controller (the confirmation signal comes 
from the watchdogs) meaning that the packet can be played. This is done using signals \emph{vokk} 
and \emph{sokk}. The confirmation is immediately forwarded to the presentation device 
using signals \emph{vpresent} and \emph{spresent}. Since the presentation device is not 
modelled, those actions are internal.

\paragraph{The watchdog timers}
The role of watchdog timers is to ensure that the time between presentations of 
subsequent media packets is within certain time bounds. We discuss the video watchdog. The
timing requirement is that consecutive video packets are played between 35 ms and 45 ms.
Initially the watchdog waits for the first packet to arrive (signal \emph{vok}) and sends 
immediately the confirmation to the video manager using signal \emph{vokk}. We ensure that
no time passes between \emph{vok} and \emph{vokk} signals. The signals \emph{vok} and
\emph{vokk} constitute in a way a complex signal that allows synchronisation between 
\emph{VideoSynch}, \emph{VideoMgr} and \emph{VideoWdg}. After presenting the first 
packet the time is measured until the next packet arrives. In order to ensure proper 
timing of the presentation of the packet, the transition leaving location vw3 is 
guarded by $35\leq$t4$\leq 45$. If \emph{vok} does not occur before 45 ms passes, 
\emph{vlate} error is given.

The \emph{SoundWdg} is slightly more complicated because we must ensure that sound 
packets are played exactly every 30 ms. Similarly to the video watchdog it waits for the 
confirmation (signal \emph{sok) }that the first packet should be presented and relays the 
signal to the \emph{SoundMgr} using the signal \emph{sokk} without time delay. After that 
the clock t3 is used to measure the time between consecutive presentations of the sound 
packets. To ensure that exactly 30 ms passes between sound packets, \emph{UrgMon} is used 
and signal \emph{sok} is marked as urgent. If a sound packet is not available in 30 ms the 
\emph{slate} error is generated.

\paragraph{The synchroniser}
The role of a \emph{Synch} is to initialise the other automata. Depending on whether a 
sound or a video packet arrives first, automata can be initialised in two ways. If signal 
\emph{vready} or \emph{sready} arrives then we confirm that the packet can be presented 
(signal \emph{vok} or \emph{sok}) and initialise \emph{SoundClock} (signal \emph{std} or 
\emph{sti}) then initialise \emph{VideoSych} (signal \emph{sv1} or \emph{ss1}) and at last 
we initialise \emph{SoundSynch} (signals \emph{sv0} or \emph{ss0}). Note that all 
locations except sy1 are committed to ensure that initialisation is done immediately.

\paragraph{The sound clock}
The \emph{SoundClock} is a discrete clock that ticks every millisecond. It is started 
at the moment the first sound packet arrives. It can be initialised by signal \emph{sti} 
if the sound packer is first or by combination of \emph{std} and \emph{sclock} signals 
if video packet arrives first.

The clock is used to compute the skew between sound and video streams. The skew is stored in 
\emph{vmins} variable. Every time the clock ticks it is decreased by one.

\paragraph{The sound synchroniser}

The \emph{SoundSynch} can be initialised it two ways. If a sound packet is first it receives 
\emph{ss1} signal and starts the repeating behaviour immediately. If a video packet is first 
then it checks if there is a synchronisation error - it can happen only when the first sound 
packet does not arrive within 15 ms after the first video packet. This is the requirement 
of the lip synchronisation. After the fist sound packet arrives it initialises 
\emph{SoundClock} through the \emph{sclock} signal and starts the repeating behaviour

The repeating behaviour is very simple. If it receives signal \emph{sready} that the 
sound packet has arrived, it send a signal \emph{sok} indicating that it can be presented.

\paragraph{The video synchroniser}
The \emph{VideoSynch} is quite complex. Similarly as \emph{SoundSynch} it can be 
initialised in two ways. If the video packet arrives first it goes immediately to the
repeating behaviour through signal \emph{sv1}. If sound packet arrives first, it checks if 
a video packet is received within 150 milliseconds but not earlier than 15 milliseconds 
from the sound packet. If more than 150 milliseconds have passed then \emph{vsynch\_error} 
is generated.

In the repeating behaviour, \emph{VideoSynch} checks if there is too much skew between 
sound and video packets. After receiving \emph{vready} signal, it checks the lip 
synchronisation requirement immediately (the t1$\leq 0$ invariant). Now we have three 
possibilities:

\begin{itemize}
	\item The video presentation is more than 150 milliseconds later than the sound 
        presentation. This is true when \emph{vmins} is less than $-150$. In such case 
        \emph{vsynch\_error} is generated.
	\item The video is more than 15 ms early with respect to the sound packet. 
        This is the case if \emph{vmins}$>15$. In such case the presentation of the video 
        frames are postponed. We enter a state where we are forced to wait one millisecond. 
        After that we check the synchronisation requirement again.
	\item The video and the sound packets are sufficiently synchronised. In such case 
        we send a signal \emph{vok} to present a video packet and we update the 
        \emph{vmins} variable.
\end{itemize}

\paragraph{The media streams}
The informal specification of the protocol does not make any assumptions about the 
streams. Several possible streams are modelled and are further described in 
Section \ref{sec:verification}.

\section{Verification}
\label{sec:verification}

\subsection{Verified properties} 

\begin{figure}[htbp]%
\centering
\includegraphics[scale=0.7]{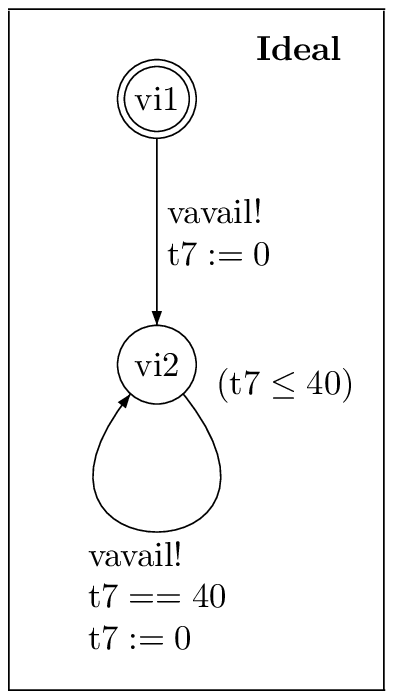}
\includegraphics[scale=0.7]{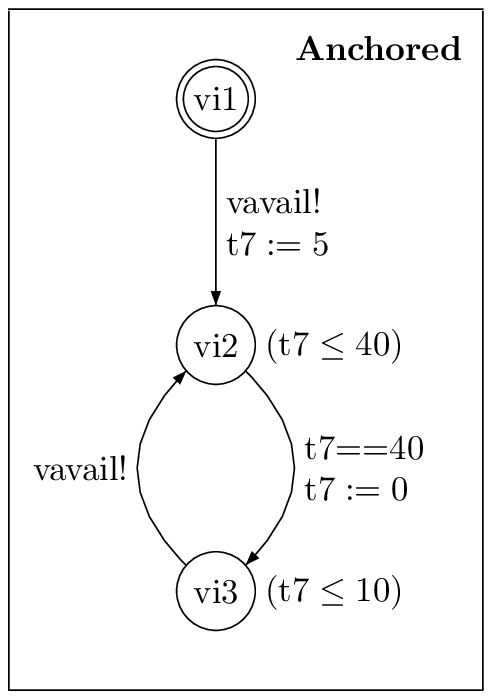}
\includegraphics[scale=0.7]{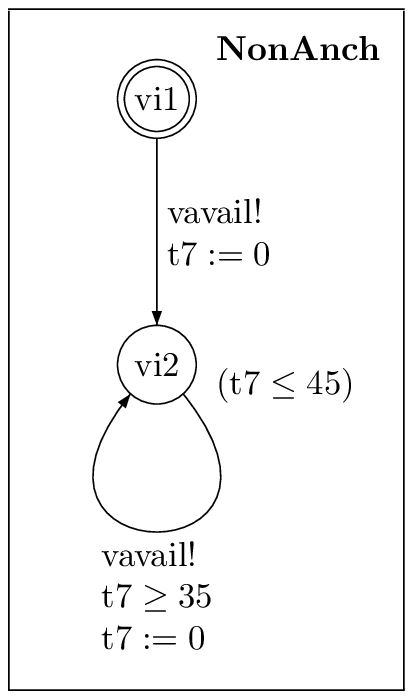}
\caption{Three variations of video stream}
\label{fig:videoStream}
\end{figure}
We have followed \cite{Bowman98} and we assumed that the sound stream is ideal and arrives 
every 30 ms. The perturbations may affect the video stream. There are three possible video 
streams that are investigated:
\begin{itemize}
	\item An \emph{ideal} video stream that delivers a video frame every $40$ ms. 
	\item A video stream with \emph{anchored jitter} that have a rate of $40$ ms and
              a variation of $\pm 5$ ms.
	\item A video stream with \emph{non-anchored jitter} where the variability between 
              each two consequent frames is minimally 35 ms and maximally 45 ms.
\end{itemize}
Figure \ref{fig:videoStream} shows automata representing different variations of the video 
stream.

Another variation that was investigated is the initial delay of video and sound streams. 
The first option is that the starting time of streams is left unspecified; the other 
possibility is that both streams start at the same time.

The verification is done using error location reachability. Each error location 
reachability was done using normal and robust semantics. The reachability properties 
are all of the form:
\[
E\Diamond A.l\;\;and\;\;not\left(B_1.l_1\;\;or\;\dots\;B_n.l_n\right)
\] 
The answer to such a query will be positive if there exists a path in timed automata 
network which will eventually reach location $l$ in $A$ but all locations $l_i$ in $B_i$
will be avoided. The location $l$ will be the error location we are checking.
The second part is to ensure that timed automata network did not reach another error
location as this might have caused another error location to be reachable.
The following error location have been modelled and checked for reachability:
\begin{itemize}
	\item Initial sound synchronisation error in the \emph{SoundSynch} (location s07)
	\item Initial video synchronisation error in the \emph{VideoSynch} (location v06)
	\item Video synchronisation error in the \emph{VideoSynch} (location v07)
	\item Video late error in the \emph{VideoWdg} (location vw5)
	\item Sound late error in the \emph{SoundWdg} (location sw5)
\end{itemize}

\subsection{Results} 
We have implemented the algorithm for the robust semantics in a prototype tool based
on UPPAAL $4.1.1$. The normal semantics reachability analysis is done using UPPAAL
$4.1.1$. Our implementation at best can be as good as a depth first search for the 
normal reachability. Thus all the results presented here are run using depth first 
search. Experiments were performed on a PC with an AMD 1.2 GHz processor with 768MB 
of RAM. In \cite{Bowman98} the state space was reduced by marking the error locations 
as committed. We have not done this optimisation. 

\begin{table}[htbp]
	\centering
     \begin{tabular}{|r|c|c|c|c|c|c|} 
     \hline
     Error location & \multicolumn{2}{|c|}{Ideal}%
     &\multicolumn{2}{|c|}{Anchored}%
     &\multicolumn{2}{|c|}{Non-anchored}\\ \hline
Init Sound Synch(s07) & T & 0.5 & T &0.1& T &0.2\\ \hline
Init Video Synch(v06) & T & 1.5 & T &56.7& T &55.7\\ \hline
Video Synch (v07)     & F &  3.3& T &57.9& T &26.5\\ \hline
Video Late (vw5)      & F &  3.3& T &0.2& F &55.5\\ \hline
Sound Late (sw5)      & F &  3.2& F &61.8& F &57.2\\ \hline
Init Sound Synch(s07*) & T & 0.1 & T &0.2& T &0.2\\ \hline
Init Video Synch(v06*) & T & 7.4 & T    &4613.3& T &1260.1\\ \hline
Video Synch (v07*)     & T &  3378.2& T &3168.4& T &674.9\\ \hline
Video Late (vw5*)      & F &  5636.4& T &7.5& F &5834.5\\ \hline
Sound Late (sw5*)      & F &  5378.2& F &5591.2& F &5724.4\\ \hline
     \end{tabular}
	\caption{Verification results for streams with possible initial delay for both 
           normal and robust semantics (marked with *)}
	\label{tab:res1}
\end{table}
Table \ref{tab:res1} gives the results of the verification of the lip-synchronisation 
protocol for the various reachability properties. In the leftmost column we have a type of 
error that can occur. In the case of the error location being not reachable we mark it with F and 
if the error location is reachable we put T. In the second column we have a verification time 
given in seconds.

We can see that for all kinds of video streams, the initial sound and video synchronisation 
errors can occur.
This can be explained by the fact that video or sound stream can postpone sending packet. 
This allows for the gap between sound and video packet to be arbitrarily long and reaching 
error location.

The anchored and non-anchored video streams can encounter video synchronisation error and 
can reach location vw07. In both cases it is enough to wait as long as possible but 
avoiding initial video synchronisation error and then the gap between sound and video 
packet can be enlarged due to allowed jitter.

Only in the case of the video stream with anchored jitter video frames can be late. In 
anchored jitter maximal gap between two consecutive packets is $50$ ms. This is $5$ ms more 
than allowed gap thus video frames can be late.

\begin{table}[htbp]
	\centering
     \begin{tabular}{|r|c|c|c|c|c|c|} 
     \hline
     Error location & \multicolumn{2}{|c|}{Ideal}%
     &\multicolumn{2}{|c|}{Anchored}%
     &\multicolumn{2}{|c|}{Non-anchored}\\ \hline
Init Sound Synch(s07) & F & 0.2 & F &0.4& F &86.1\\ \hline
Init Video Synch(v06) & F & 0.2 & F &0.5& F &85.1\\ \hline
Video Synch (v07)     & F &  0.2& F &0.5& T &36.8\\ \hline
Video Late (vw5)      & F &  0.2& T &0.2& F &83.1\\ \hline
Sound Late (sw5)      & F &  0.2& F &0.5& F &81.8\\ \hline
Init Sound Synch(s07*) & F & 81.8 & F &153.2& F &178.4\\ \hline
Init Video Synch(v06*) & F & 391.1 & F &397.2& F &357.9\\ \hline
Video Synch (v07*)     & T &  275.4 & T &293.7& T &244.2\\ \hline
Video Late (vw5*)      & F &  394.4& T &9.2& F &385.4\\ \hline
Sound Late (sw5*)      & F &  391.5& F &385.3& F &401.2\\ \hline
     \end{tabular}
	\caption{Verification results for streams without initial delay for both normal 
           and robust semantics (marked with *)}
	\label{tab:res2}
\end{table}     

The results where both streams are forced to start at the same time are shown in Table 
\ref{tab:res2}. The presentation format is the same as in Table \ref{tab:res2}. In such a
situation the initial synchronisation errors are not reachable. In the case of an ideal video 
stream we do not encounter any errors. In the case of an anchored video stream again the 
video can be late. The reason is the same as previously. The non-anchored video stream 
can lead to out of video synchronisation error. This is because the gap can accumulate 
over time.

The second part of Table \ref{tab:res1} and Table \ref{tab:res2} shows the result for 
the robust semantics. They are marked with * next to the location name. It is worth 
mentioning here that if some location is reachable in the normal semantics then it is 
reachable in the robust semantics. The main difference between normal and robust 
semantics is that a video synchronisation error is always possible for all kinds of video 
stream, with or without allowed initial delay.

We try to explain this for the case of ideal video stream without initial delay, as all 
other cases are less restrictive. The variable vmins is decreased every millisecond. 
The timing is provided by the clock t$5$. In the case of an ideal video stream, every 40 ms 
(ensured by clock t$7$) a video packet is sent. If vmins is small enough, so we do not 
have to enter location v04 and vmins is increased by 40. Thus over time vmins oscillates 
around the same base value. Now assume that clocks t$5$ and t$7$ desynchronise by 
$\eps$ every millisecond. For sufficiently many cycles the base value over 
which vmins oscillates can be changed up or down. If vmins is too large it will be 
remedied by visiting location v04, but no such mechanism exists when vmins is getting 
smaller. Finally we will reach v07 and video synchronisation error will occur. Other 
types of errors use the clocks that cannot accumulate the drift because their reset time 
is synchronised with the signals. So the verification results from normal and robust 
semantics are the same.

The conclusion is that if we play video long enough we are not able to guarantee that 
the protocol will not desynchronise, no matter how precise clocks we have. We are only 
able to guarantee proper lip synchronisation for a playback with limited time.

\paragraph{Deadlocks}
In \cite{Bowman98} in addition the deadlock detection is done. We do not do deadlock 
detection as current status of the theory does not allow the tool to detect deadlocks 
robustly. The main limiting factor is the fact that we do not allow the guards to be 
strict. To do the deadlock detection we need to detect when we reach state that cannot 
leave through any transition. For that we need to complement guards on the edges and that 
introduces strict inequalities. The brief manual analysis of  the specification reveals 
that new deadlocks can be reached in robust semantics. For example non-anchored video can
send video packet at 45 ms but because of slight desynchronisation clock t$4$ have value
$45+\eps$ thus edge leading to location vw4 is not enabled any more.

As a side note let us mention that in \cite{Bowman98} authors report one deadlock and 
they expected the other deadlock that should be detected by the UPPAAL, but full state 
search did not revealed it. It appears that the reason for not detecting the deadlock must 
have been the early imperfection of the UPPAAL tool, as the current version $4.1.1$ detect 
both deadlocks. 

\section{Conclusions}
\label{sec:conclusions}
We have re-verified a lip synchronisation protocol using robust
semantics. The original protocol has been previously model checked 
using UPPAAL \cite{Bowman98} and has been presented
in a number of different formalisms \cite{StefaniHH92,Regan93,AtesBSS96}.

The verification results using robust semantics are slightly different
from normal semantics. The choice of case study was done to maximise 
probability of different results so this result was anticipated. 
The robust reachability analysis allowed us to identify the problem
with the lip synchronisation protocol. For a continuous playback we
are not able to make the clocks precise enough to ensure that video 
and sound do not become desynchronised. The sound and video can stay 
synchronised only for a limited time, and this time is depending on
the precision of the clocks.

The verification of the lip synchronisation protocol gave us also the
possibility to evaluate the performance of the robust reachability algorithm.
The verification of a lip synchronisation using robust semantics takes
significantly more time than verification using normal semantics. The main reason is that the
model uses the variable vmins as a kind of discrete clock which divides the state space into many
small pieces. This forces our algorithm to add many stable zones, which is expensive.

The limitation of our algorithm is its inability to detect deadlocks. In
the case of the industrial case study this is important feature. We believe that the
ability to detect deadlocks would identify more problems - mostly connected to the way
time-out is modelled. Another limitation is that it is not possible to use strict
guards. In the context of robustness where we allow small clock drifts, we believe
that differentiating between strict and non-strict guards is not an essential feature.
Unfortunately this feature is needed for deadlock detection. This will be an important direction of
our future work. At the moment only reachability properties can be analysed. Future
research will investigate the extension of the algorithm with the
possibility to check liveness properties.
\bibliographystyle{eptcs} 

\bibliography{lsaBib}
\end{document}

%% file: macros.tex
\newcommand{\x}{x}           
\newcommand{\y}{y}           
\newcommand{\va}{v}					 
\newcommand{\q}{q}					 
\newcommand{\e}{e}           
\newcommand{\s}{s}        
\newcommand{\Sem}{\mathcal{S}}						
\newcommand{\Reach}[1]{\mathsf{Reach}(#1)}  
%
%
\newcommand{\eps}{\varepsilon} 
\newcommand{\stepE}[1]{\stackrel{#1}{\rightarrow}_{\eps}}	
\newcommand{\ReachE}[1]{\mathsf{Reach}_\eps({#1})}  
\newcommand{\StarReachE}[1]{R_{\eps\to 0}({#1})}  	
\newcommand{\II}[2]{\vbox{\hbox{#1}\hbox{#2}}}
\newcommand{\III}[3]{\vbox{\hbox{#1}\hbox{#2}\hbox{#3}}}
\newcommand{\IV}[4]{\vbox{\hbox{#1}\hbox{#2}\hbox{#3}\hbox{#4}}}
\newcommand{\IX}[9]{\vbox{\hbox{#1}\hbox{#2}\hbox{#3}\hbox{#4}\hbox{#5}\hbox{#6}\hbox{#7}\hbox{#8}\hbox{#9}}}